# CO Induced Adatom Sintering in a Model Catalyst: Pd/Fe$_3$O$_4$

Gareth S. Parkinson[1]*, Zbynek Novotny[1], Giacomo Argentero[1], Michael Schmid[1], Jiří Pavelec[1], Rukan Kosak[2], Peter Blaha[2], Ulrike Diebold[1]*

[1]Institute of Applied Physics, Vienna University of Technology, Wiedner Hauptstrasse 8-10/134, 1040 Vienna, Austria

[2]Institute of Materials Chemistry, Vienna University of Technology, Getreidemarkt 9/165-TC, 1060 Vienna

*Correspondence to: parkinson@iap.tuwien.ac.at; diebold@iap.tuwien.ac.at

**Introductory paragraph:** The coarsening of catalytically-active metal clusters is often accelerated by the presence of gases through the formation of mobile intermediates[1-4], though the exact mechanism through which this happens is often subject to debate[1,5-10]. We use scanning tunneling microscopy (STM) to follow the CO induced coalescence of Pd adatoms supported on the Fe$_3$O$_4$(001) surface at room temperature. We show that highly-mobile Pd-carbonyl species, formed via the so-called "skyhook" effect[11], are temporarily trapped at other Pd adatoms. Once these reach a critical density, clusters nucleate; subsequent coarsening occurs through cluster diffusion and coalescence. While CO increases the mobility in the Pd/Fe$_3$O$_4$ system, surface hydroxyls have the opposite effect. Pd atoms transported to surface OH groups are no longer susceptible to the skyhook effect and remain isolated. Following the evolution from well-dispersed metal adatoms into clusters, atom-by-atom, allows identification of the key processes that underlie gas-induced mass transport.

Gas-enhanced mass transport at surfaces is particularly important in heterogeneous catalysis, where the growth of supported precious metal nanoparticles during reactions, known as sintering, leads to catalyst deactivation [6,12,13]. Sintering occurs via two mechanisms; Ostwald ripening, where atoms detach from smaller clusters and diffuse to large ones, and cluster diffusion and coalescence. Adsorbed gas molecules can impact these processes by (i) aiding in the break up of clusters [5-7], (ii) modifying the potential energy landscape of the support [1,14-16], and (iii) formation of mobile molecule-metal intermediates [1,5,7-10].

The basic tenet, that adsorption can enhance the mobility of surface atoms, was demonstrated by Horch et al. [11] for the H-Pt system with scanning tunneling microscopy (STM). Upon H adsorption, the mobility of surface Pt atoms increased dramatically; this phenomenon was termed the "skyhook effect." Despite this early success, no model catalyst system has been found in which a well-defined initial state evolves on an observable timescale, allowing individual species to be followed and the conduit of mass transfer determined. Instead, experimental studies typically focus either on observation of the evolving cluster size distribution in situ [8,9,17-21], or infer the mechanism from a post mortem analysis of the cluster size distribution. The true nature of the diffusing species is unknown in many cases[1].

We follow the CO-induced coalescence of isolated Pd adatoms at the $Fe_3O_4$(001) surface atom-by-atom using time-lapse STM. By tracing the evolution of the system all the way from single atoms to larger Pd clusters, we identify the CO-induced onset of mobility, the mechanism for cluster nucleation and coarsening, and the role of surface hydroxyls.

The work relies on the adatom templating property of the $Fe_3O_4$(001) support [22]. STM images (Fig. 1A) show rows of octahedrally coordinated $Fe_{oct}$ atoms along <110> directions in a simple truncation of the bulk inverse spinel structure (Fig. 1B). The undulating appearance of the $Fe_{oct}$ rows results from small lateral relaxations (≈ 0.1 Å) [23], which create a ($\sqrt{2}\times\sqrt{2}$)R45° symmetry (cyan square). Typically, 0.05 monolayers (1 ML = 1 adsorbate per ($\sqrt{2}\times\sqrt{2}$)R45° unit cell, $1.42\times10^{14}$ atoms/cm$^2$) of hydroxyl groups are observed (Fig. 1A), which result from the dissociative adsorption of water at the oxygen vacancies created during sputter/anneal cycles [24]. The hydroxyl modifies the density of states of a nearby $Fe_{oct}$ pair, which then appears brighter in STM [25], see Fig. 1B.

In addition to this lattice distortion, density functional theory-based (DFT+U) calculations predict long-range order amongst $Fe^{3+}$ and $Fe^{2+}$ cations in subsurface layers [26]. Adatoms are expected to adsorb where tetrahedrally-coordinated $Fe_{tet}$ atoms would reside if the bulk structure were continued. However, Fe, Au, Pd and H atoms strongly prefer only one of two closely related sites [22,24,25,27], at positions where neighboring $Fe_{oct}$ rows relax toward one another. This adsorption template stabilizes isolated Au adatoms (minimum separation 8.4 Å) up to sample temperatures of 670 K [22].

According to our GGA+U calculations Pd atoms adsorb in a symmetric configuration in such a tetrahedral site (Fig. 1C) with a high adsorption energy (2.2 eV). The Pd atom has (formal) 2+ charge, a magnetic moment of 0.47 $\mu_B$, and the same polarization as the $Fe_{tet}$ atoms. In experiment, deposition of 0.2 ML Pd at RT leads to uniformly bright protrusions between the characteristic $Fe_{oct}$ rows of the substrate (Fig. 2A). These are single Pd atoms. As found for other adsorbates on $Fe_3O_4$(001), the Pd atoms occupy preferentially one of the two available sites (the white squares in the schematics in Fig. 2A).

Despite the strong bonding of the Pd atoms, some mobility is observed at RT. The STM images in Fig. 2A-D show selected frames from a series of images recorded at RT over more than 5 hours in a background pressure of $6\times10^{-11}$ mbar (Movie S3). Note that the area shown in Fig. 2 is a small cutout from the much larger area (50×50 nm$^2$) in the STM movie. In the region of interest, the first 50 images of the STM movie are identical to Figure 2A, but then in image 51 (Fig. 2B), Pd(2) is replaced by a bright feature, while Pd(1) disappears. Since this is the only change in the vicinity, the new protrusion must contain both Pd(1) and Pd(2). Its fuzzy appearance is characteristic of a species that moves during scanning; an indication of weak binding to the surface. A few frames later (frame 55, Fig. 2C) the feature jumps to Pd(3), leaving Pd(2) behind. Figure 2D shows a similar event whereby the feature hops from Pd(3) to Pd(4). Thus only one adatom becomes mobile, and is temporarily trapped via interaction with another Pd adatom before continuing across the surface. Since events such as those observed in Fig. 2 are rare, and separated by tens of nm, we can track the motion of individual Pd atoms on the surface.

In addition to pausing at other Pd adatoms, the mobile Pd's also interact with surface hydroxyl groups. Fig. 3 A-D shows consecutive STM images selected from another STM movie for a sample similar to that in Fig. 2. Frame (A) shows isolated Pd atoms, hydroxyls, and a brighter, solid protrusion with an apparent height of 270 pm (red cross). Between Figs. 3(A) and (B), Pd(1) becomes mobile and diffuses to Pd(2), creating the fuzzy feature in (B); a second fuzzy feature has been formed in (C). In Fig. 3(D), however, the mobile adatom has diffused away from Pd(2), and has formed a new species at the position of a hydroxyl group. This new species has the same apparent height as the one marked by a red cross in Fig. 3(A). It does not move under the STM tip (i.e., does not appear fuzzy), and stays put after formation. Again, as only one diffusion event occurs over a wide region, it is clear that one Pd atom and one H atom are contained within this newly-formed species.

Most of the isolated Pd adatoms at the $Fe_3O_4$(001) surface at RT are completely immobile over several hours in UHV, except for the occasional Pd that spontaneously becomes mobile and rapidly diffuses over large distances on the surface. This behavior suggests a 'skyhook' is at work [11], i.e., adsorption of a molecule from the gas phase weakens the bond of a surface atom, allowing facile diffusion of the adatom-molecule intermediate.

We show that CO, the major constituent of the UHV residual gas, is responsible for mobility in the Pd/$Fe_3O_4$(001) system, via formation of a Pd carbonyl (CO-Pd) intermediate. In GGA+U calculations, when a CO molecule approaches the Pd (Fig. 1D), one Pd-O bond is broken and the other one weakens, increasing in length to 2.05 Å. The CO attaches strongly to the Pd, but weakens its internal bonding as the C-O distance increases to 1.16 Å. The Pd atom loses its magnetic moment, acquiring a neutral charge state. In experiment, exposure to CO leads to a coarsening of the Pd on the $Fe_3O_4$(001) surface. Fig. S1 shows two large-scale STM images recorded before and after exposure of a 0.2 ML Pd/$Fe_3O_4$(001) surface to $1.33\times10^{-4}$ mbar×sec CO.

Prior to CO exposure, isolated Pd atoms, hydroxyl groups and H-Pt features (red ×) are observed. Following CO exposure no isolated Pd adatoms remain. Instead, large clusters of varying size have formed, while the number of H-Pd features has increased by a factor of 4. We can conclude that CO exposure clearly leads to cluster formation and growth in the Pd/Fe$_3$O$_4$(001) system. Interestingly, Pd atoms are significantly more stable against CO when they are stabilized by a surface hydroxyl.

To follow the CO induced cluster formation, we recorded an STM movie during which the background CO pressure was purposely raised to 5×10$^{-10}$ mbar (movie S4). As before, the initial Pd adatom coverage was 0.2 ML. In Fig. 4 we show selected frames from a small part of the area imaged by STM. Fig. 4A, acquired immediately prior to CO exposure, shows isolated Pd's, hydroxyls and one H-Pd species. After 15 minutes of CO exposure (Fig. 4B), seven Pd's have become mobile, five of which reside at Pd adatoms (bright fuzzy features), whilst two have formed additional H-Pd. Three of the mobile adatoms are in close proximity to each other (left side of Fig. 4B). In the next frame (Fig. 4C) a cluster has formed from three mobile adatoms, their host Pd adatoms, and four of the nearby adatoms that are now missing. In Fig. 4D, the large cluster has diffused 6 nm, picking up a Pd carbonyl and its host atom along the way, before merging with a H-Pd species. Tracking each adatom over the course of the movie, we determine the size of the final, large cluster to be between 15 and 19 adatoms. The uncertainty stems from some atoms close to step edges, whose final destination is harder to determine unambiguously.

While the crucial first step in the ripening of the Pd/Fe$_3$O$_4$(001) system is the creation of mobile CO-Pd species, this alone does not lead to cluster formation; the interaction between a Pd carbonyl and a Pd adatom is not strong enough to form a stable dimer that would act as a nucleation site. Formation of a stable cluster requires the interaction of multiple mobile Pd carbonyl species; as CO is added to the Pd/$Fe_3O_4$ system their number increases and the adatom density decreases (Fig. S2). The nucleation process is also evident in Fig. 4 (B-C), where three Pd-CO's are in close proximity immediately prior to the appearance of a large cluster, which also incorporates several nearby Pd's. Since Pd adatoms do not diffuse without the aid of CO, the cluster nucleus itself must diffuse rapidly on formation, collecting the Pd. This process stops once a stable size is reached. Further growth also occurs through cluster migration and coalescence, for example, the ≈15 atom cluster shown in Fig. 4C diffuses 6 nm to merge with a H-Pd species in Fig. 4D.

While an analysis of the final result – a cluster anchored at a surface H species – could suggest heterogeneous nucleation, our atom-by-atom tracking clearly shows that, in fact, homogeneous nucleation occurs, i.e., a critical density of mobile Pd-CO species causes the collapse of dispersion in this system. Further coarsening occurs via cluster diffusion. Both results are surprising for metals on oxide surfaces, where heterogeneous nucleation at defects and Ostwald ripening are generally assumed to cause cluster nucleation and growth.[9] While CO induces mobility, H has the opposite effect. A mobile Pd-CO that encounters a hydroxyl while diffusing across the surface forms a stable H-Pd complex, seemingly impervious to the CO skyhook. This process, and the associated lack of diffusion, may explain reports of highly disperse Au nanoparticles [15] and improved layer-by-layer growth [14] at hydroxylated oxide surfaces.

Key to our study are the isolated, stable adatoms at the $Fe_3O_4$(001) support. Each individual atom can be followed, and the history and size of a cluster can be determined with a high degree of accuracy. As single adatom stability appears to be an inherent property of $Fe_3O_4$(001) [22,25,27], this system promises a treasure trove of experimental data for understanding basic mechanisms in catalysis; in principle, any metal / gas combination can be studied, and fundamental mechanisms of mass transport elucidated in a direct way.

**Methods**

The experiments were performed using a synthetic $Fe_3O_4$ single crystal grown via the floating zone method by Prof. Mao and coworkers at Tulane University, USA. The clean surface was prepared in an ultrahigh vacuum (UHV) vessel (base pressure below $10^{-10}$ mbar) by cycles of $Ar^+$ sputtering (1 keV, 1.5 μA, 10 minutes), UHV annealing (920 K, 15 minutes), and annealing in $O_2$ (920 K, $p_{O2}$ = $6\times10^{-7}$ mbar, 30 minutes). Pd (99.99% purity) was deposited from rod with a carefully degassed liquid-$N_2$ cooled e-beam evaporator (Omicron EFM3). In this paper, the Pd coverage is defined in terms of the narrow sites in the surface layer, i.e. 1 monolayer (ML) Pd = 1 Pd atom per ($\sqrt{2}\times\sqrt{2}$)R45° surface unit cell ($1.42\times10^{14}$ atoms/cm$^2$). The deposition rate was 1 ML/min as calibrated using a water-cooled quartz crystal microbalance. During deposition a retarding voltage of +1.5 kV was applied to the cylindrical electrode in the orifice of the evaporator (flux monitor) to repel positively charged ions that are produced in the EFM3. STM measurements were performed in a separate analysis chamber (base pressure $6\times10^{-11}$ mbar) using a customized Omicron μ-STM operated in constant current mode at room temperature with an electrochemically etched W tip. CO (purity 99.997 %) was dosed as received via a high precision UHV leak valve. The CO partial pressure and purity of the CO gas were monitored using a Balzers QMG 125 mass spectrometer with secondary electron multiplier.

The density functional theory (DFT) based simulations utilize the augmented plane wave + local orbitals (APW+lo) method as embodied in the WIEN2k code[28]. We employ the generalized gradient approximation of Perdew et al.[29] and treat the correlated Fe-3d electrons with a Hubbard –U correction (GGA+U) and the double counting correction in the fully localized limit[30]. Specifically we employ an effective U value of 3.8 eV for the Fe-3d states. In the APW+lo method, space is divided into spheres around the atoms and an interstitial region. We use atomic sphere radii of 1.86, 1.6, 2.1, 0.9 and 1.2 bohr for Fe, O, Pd, C and O (of CO), respectively. The plane wave cutoff is set to 20 Ry and the 2D Brillouin zone is sampled with a 2x2x1 k-mesh. We use a Fermi broadening of 0.08 eV.

The $Fe_{oct}$-terminated $Fe_3O_4$ (001) surface is modeled using a symmetric slab with 17 layers and a 2x2 supercell (248 Fe and O atoms). In agreement with previous calculations (Ref.30,33) we find charge ordering of the sub-surface $Fe_{oct}$ atoms with a characteristic dimerization leading to the well-known (√2x√2)R45° reconstruction of the clean surface. The larger supercell used in the present calculations allows for a better separation of the Pd (CO) ad atoms.

**Acknowledgements**

This material is based upon work supported as part of the Centre for Atomic-Level Catalyst Design, an Energy Frontier Research Centre funded by the U.S. Department of Energy, Office of Science, Office of Basic Energy Sciences under Award Number #DE-SC0001058. The authors acknowledge Prof. Z. Mao and T.J. Liu (Tulane University) for the synthetic sample used in this work. GSP acknowledges support from the Austrian Science Fund project number P24925-N20. RK and JP acknowledge stipends from the TU Vienna doctoral college CATMAT.

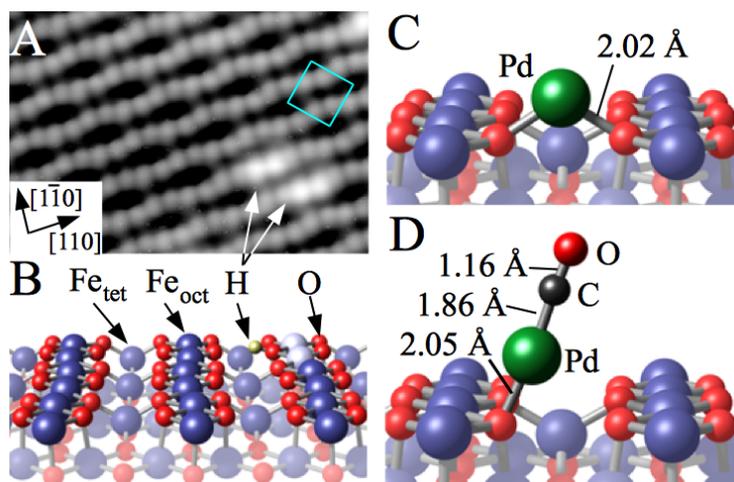

Fig. 1: The $Fe_3O_4(001)$ surface. (A) Scanning tunneling microscopy image (6.4×4.4 $nm^2$, $V_{sample}$ = +1.4 V, 0.1 nA) of the freshly prepared surface at RT. The ($\sqrt{2}\times\sqrt{2}$)R45° unit cell of the distorted B-layer termination is indicated by a cyan square. Surface hydroxyl groups, formed through dissociation of water at O vacancies, appear as double bright protrusions because the H atom modifies the density of states of a pair of neighboring surface $Fe_{oct}$ atoms. (B) Schematic model of the distorted B-layer termination of $Fe_3O_4(001)$. Only surface $Fe_{oct}$ atoms appear in STM images. One bright pair of $Fe_{oct}$ atoms are drawn along with the associated hydrogen atom. (C, D) Equilibrium geometry around the adsorbed Pd atom on the $Fe_3O_4$ (001) surface from GGA+U calculations. (C) A single adsorbed Pd atom; (D) a CO molecule attached to the Pd atom.

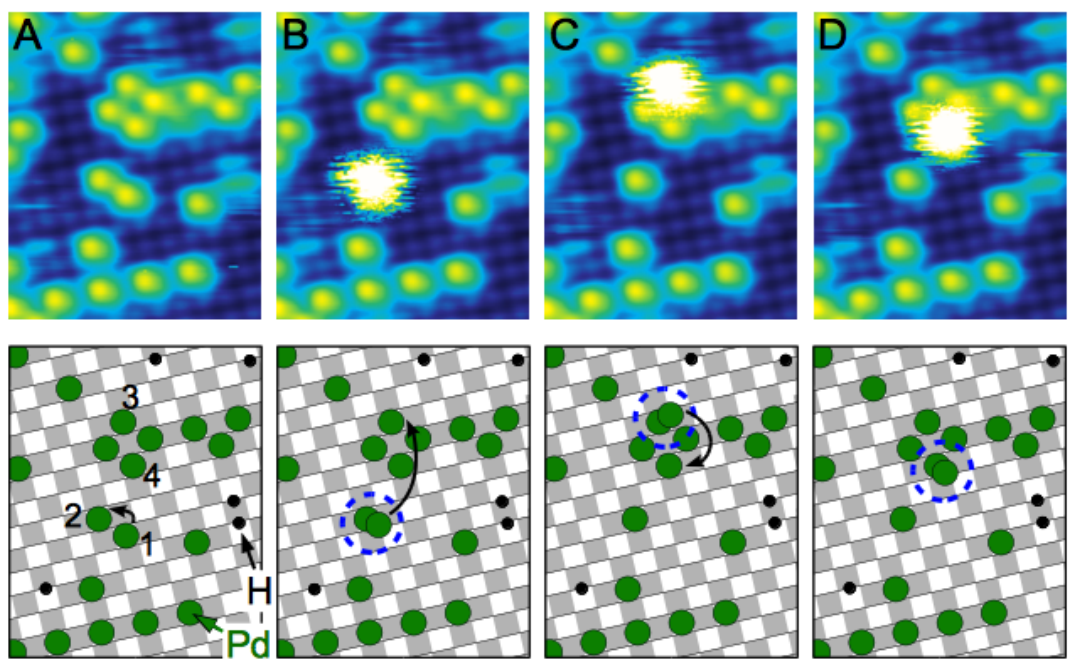

Fig. 2: CO induced mobility of one Pd adatom. STM images (6.5×8.5 nm$^2$, +1 V, 0.2 nA) acquired following deposition of 0.2 ML Pd on the $Fe_3O_4$(001) surface at room temperature. The four images (A-D) are selected from a longer STM movie acquired over the same sample area (see Supplement). The schematic model below each image shows the motion of the atom. (A) Adatom 1 hops to adatom 2 forming a bright fuzzy feature. (B-D) The atom hops to atom 3 and then to 4, leaving behind its host adatom in each case.

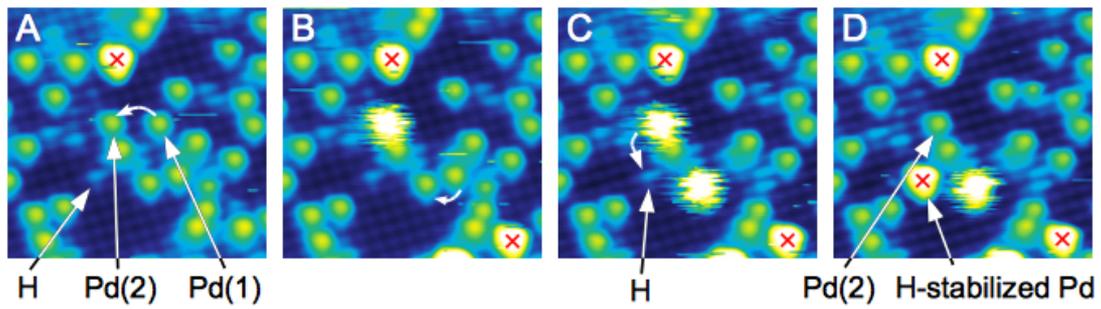

Fig. 3: Formation of a stable Pd adatom at a surface hydroxyl. Consecutive STM images (10×10 nm$^2$, +1 V, 0.2 nA) acquired over the same sample area following the deposition of 0.2 ML Pd at RT. Adatom Pd(1) hops to Pd(2), forming a bright fuzzy feature, before hopping to a hydroxyl group, where it converts into a bright, solid feature. Each bright, solid feature (marked with a red ×) contains just one single Pd atom at the position of a surface hydroxyl.

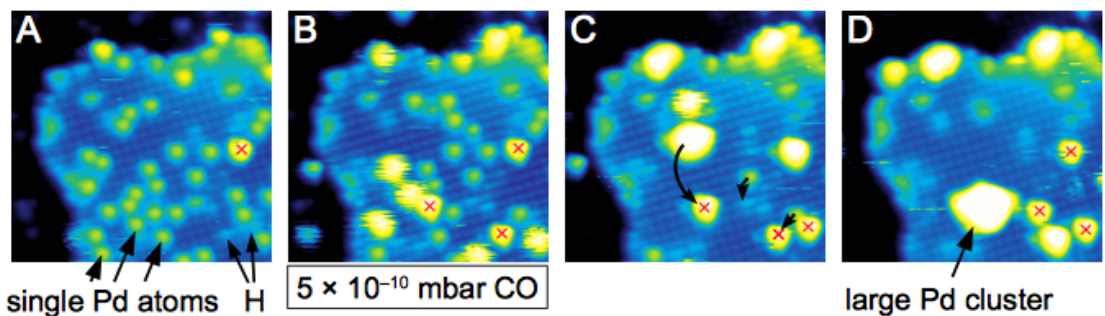

Figure 4: The CO induced formation of a large Pd cluster. Four STM images (14×14 nm$^2$, +1 V, 0.2 nA) selected from a 36 frame STM movie (duration 1 hour 50 minutes, see Supplement Movie S4) following the deposition of 0.2 ML Pd at RT. Initially (A), isolated Pd atoms are present, together with hydroxyl groups and one H-Pd (red X). After 3 frames the background pressure of CO is raised to 5×10$^{-10}$ mbar. 30 minutes later (frame B), several mobile 'fuzzy' atoms, trapped at other Pd atoms have formed. Shortly afterward (C), three mobile atoms and four adatoms have formed a large cluster. 25 minutes later (D), the cluster has captured another mobile adatom, and diffused to merge with a H-Pd species.